\documentstyle[11pt,epsf,rotate]{article}
\input{psfig}

\newcommand{\IM}{{\rm Im}}
\newcommand{\vcb}{|V_{cb}|}
\newcommand{\vtd}{|V_{td}|}
\newcommand{\vub}{|V_{ub}/V_{cb}|}

\def\R1{\varepsilon_1}
\def\E8{\varepsilon_8}

\def\epe{\varepsilon'/\varepsilon}
\def\as{\alpha_s}

\newcommand{\mt}{m_{\rm t}}
\newcommand{\mtb}{\overline{m}_{\rm t}}

\newcommand{\mc}{m_{\rm c}}
\newcommand{\ms}{m_{\rm s}}

\newcommand{\mb}{m_{\rm b}}
\newcommand{\mw}{M_{\rm W}}

\newcommand{\gev}{\, {\rm GeV}}
\newcommand{\mev}{\, {\rm MeV}}

\newcommand{\Lms}{\Lambda_{\overline{\rm MS}}}

\newcommand{\bea}{\begin{eqnarray}}
\newcommand{\eea}{\end{eqnarray}}
\newcommand{\bd}{\begin{displaymath}}
\newcommand{\ed}{\end{displaymath}}

\newcommand{\beq}{\begin{equation}}
\newcommand{\eeq}{\end{equation}}
\newcommand{\be}{\begin{equation}}
\newcommand{\ee}{\end{equation}}
\newcommand{\bi}{\begin{itemize}}
\newcommand{\ei}{\end{itemize}}
\newcommand{\ord}{{\cal O}}

\def\kpn{K^+\rightarrow\pi^+\nu\bar\nu}
\def\klpn{K_{\rm L}\rightarrow\pi^0\nu\bar\nu}

\def\aspi{\frac{\as}{4\pi}}

\begin{document}
\thispagestyle{empty}
\begin{flushright}
 TUM-HEP-343/99 \\
 hep-ph/9901409 \\
January 1999
\end{flushright}
\vskip1truecm
\centerline{\Large\bf  Operator Product Expansion, Renormalization Group}
\centerline{\Large\bf and}
\centerline{\Large\bf  Weak Decays
   \footnote{\noindent
   Dedicated to the 70th birthday of Wolfhart Zimmermann.\\
To appear in {\it Recent Developments in Quantum
Field Theory}, Springer Verlag, eds. P. Breitenlohner, 
D. Maison and J. Wess.   }}
\vskip1truecm
\centerline{\large\bf Andrzej J. Buras}
\bigskip
\centerline{\sl Technische Universit{\"a}t M{\"u}nchen}
\centerline{\sl Physik Department} 
\centerline{\sl D-85748 Garching, Germany}
\vskip1truecm
\centerline{\bf Abstract}
A non-technical  description of the Operator
Product Expansion and Renormalization Group techniques as applied
to weak decays of mesons is presented.
We use this opportunity to summarize briefly
the present status of the next-to-leading QCD corrections to weak
decays and their implications for
the unitarity triangle, the ratio $\epe$,  the radiative decay
$B\to X_s\gamma$, and  the rare decays 
$K^+\to\pi^+\nu\bar\nu$ and $K_L\to\pi^0\nu\bar\nu$.

\newpage

\thispagestyle{empty}

\mbox{}

\pagenumbering{arabic}

\setcounter{page}{1}

\section{Preface}

It is a great privilege and a great pleasure to give this talk 
at the symposium celebrating the 70th birthday of Wolfhart
Zimmermann.
The Operator Product Expansion \cite{OPE1} to which Wolfhart
Zimmermann contributed in such an important manner \cite{OPE2,OPE3,OPE4}
 had an
important impact on my research during the last 20 years.
I do hope very much to give another
talk on this subject in 2008 at a symposium celebrating 
Wolfhart Zimmermannïs 80th birthday. I am convinced
that OPE will play an important role in the next 10 years in the field of 
weak decays as it played already in almost 25 years
since the pioneering applications of this very powerful method by
 Gaillard and Lee \cite{MKG} and
Altarelli and Maiani \cite{ALMA}.

\section{Operator Product Expansion}
The basic starting point for any serious phenomenology of weak decays of
hadrons is the effective weak Hamiltonian which has the following generic
structure
\be\label{b1}
{\cal H}_{eff}=\frac{G_F}{\sqrt{2}}\sum_i V^i_{\rm CKM} C_i(\mu)Q_i~.
\ee
Here $G_F$ is the Fermi constant and $Q_i$ are the relevant local
operators which govern the decays in question. The Cabibbo-Kobayashi-Maskawa
factors $V^i_{CKM}$ \cite{CAB,KM} 
and the Wilson Coefficients $C_i$ \cite{OPE1} describe the 
strength with which a given operator enters the Hamiltonian.

\begin{figure}[hbt]
\vspace{0.10in}
\centerline{
\epsfysize=1.9in
\epsffile{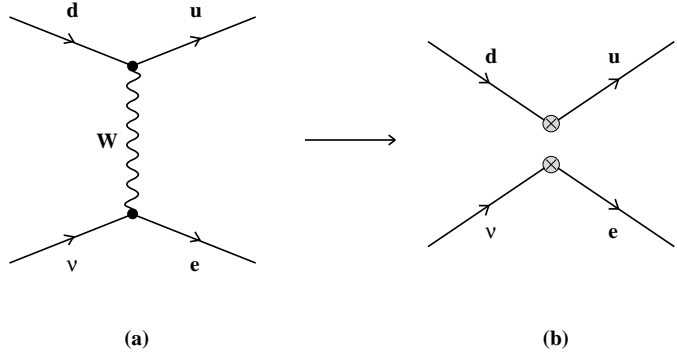}
}
\vspace{0.08in}
\caption[]{
$\beta$-decay at the quark level in the full (a) and effective (b)
theory.
\label{L:1}}
\end{figure}

In the simplest case of the $\beta$-decay, ${\cal H}_{eff}$ takes 
the familiar form
\be\label{beta}
{\cal H}^{(\beta)}_{eff}=\frac{G_F}{\sqrt{2}}
\cos\theta_c[\bar u\gamma_\mu(1-\gamma_5)d \otimes
\bar e \gamma^\mu (1-\gamma_5)\nu_e]~,
\ee
where $V_{ud}$ has been expressed in terms of the Cabibbo angle. In this
particular case the Wilson Coefficient is equal unity and the local
operator, the object between the square brackets, is given by a product 
of two $V-A$ currents. This local operator is represented by the
diagram (b) in fig. \ref{L:1}.
Equation (\ref{beta}) represents the Fermi theory for $\beta$-decays 
as formulated by Sudarshan and
Marshak \cite{SUMA} and Feynman and Gell-Mann \cite{GF} forty years ago, 
except that in (\ref{beta})
the quark language has been used and following Cabibbo a small departure of
$V_{ud}$ from unity has been incorporated. In this context the basic 
formula (\ref{b1})
can be regarded as a generalization of the Fermi Theory to include all known
quarks and leptons as well as their strong and electroweak interactions as
summarized by the Standard Model. It should be stressed that the formulation
of weak decays in terms of effective Hamiltonians is very suitable for the
inclusion of new physics effects. We will discuss this issue briefly later
on.

Now, I am aware of the fact that the formal operator language used here is
hated by experimentalists and frequently disliked by more phenomenological
minded theorists. Consequently the literature on weak decays, in particular
on B-meson decays \cite{DIAG}, 
is governed by Feynman diagram drawings with W-, Z- and top
quark exchanges, rather than by the operators in (\ref{b1}). 
In the case of the $\beta$-decay we have the diagram (a) in fig.~\ref{L:1}.
Yet such Feynman
diagrams with full W-propagators, Z-propagators and top-quark propagators
really represent the situation at very short distance scales 
$\ord ({\rm M_{W,Z}, m_t})$, whereas the
true picture of a decaying hadron with masses 
$\ord(\mb,\mc,m_K)$ is more properly described by
effective point-like vertices which are represented by the local operators
$Q_i$. The Wilson coefficients $C_i$ can then be regarded as coupling constants
associated with these effective vertices.

Thus ${\cal H}_{eff}$ in (\ref{b1}) is simply a series of effective 
vertices multiplied 
by effective coupling constants $C_i$. This series is known under the name 
of the operator product expansion (OPE) \cite{OPE1}-\cite{OPE4}, \cite{WIT}. 
Due to the interplay of electroweak 
and strong interactions the structure of the local operators (vertices) is 
much richer than in the case of the $\beta$-decay. They can be classified 
with respect to the Dirac structure, colour structure and the type of quarks 
and leptons relevant for a given decay. Of particular interest are the 
operators involving quarks only. They govern the non-leptonic decays.
To be specific let us list the operators relevant for non-leptonic
B--meson decays. They are:

{\bf Current--Current :}
\begin{equation}\label{OS1} 
Q_1 = (\bar c_{\alpha} b_{\beta})_{V-A}\;(\bar s_{\beta} c_{\alpha})_{V-A}
~~~~~~Q_2 = (\bar c b)_{V-A}\;(\bar s c)_{V-A} 
\end{equation}

{\bf QCD--Penguins :}
\begin{equation}\label{OS2}
Q_3 = (\bar s b)_{V-A}\sum_{q=u,d,s,c,b}(\bar qq)_{V-A}~~~~~~   
 Q_4 = (\bar s_{\alpha} b_{\beta})_{V-A}\sum_{q=u,d,s,c,b}(\bar q_{\beta} 
       q_{\alpha})_{V-A} 
\end{equation}
\begin{equation}\label{OS3}
 Q_5 = (\bar s b)_{V-A} \sum_{q=u,d,s,c,b}(\bar qq)_{V+A}~~~~~  
 Q_6 = (\bar s_{\alpha} b_{\beta})_{V-A}\sum_{q=u,d,s,c,b}
       (\bar q_{\beta} q_{\alpha})_{V+A} 
\end{equation}

\begin{figure}[hbt]
\centerline{
\epsfysize=4.5in
\epsffile{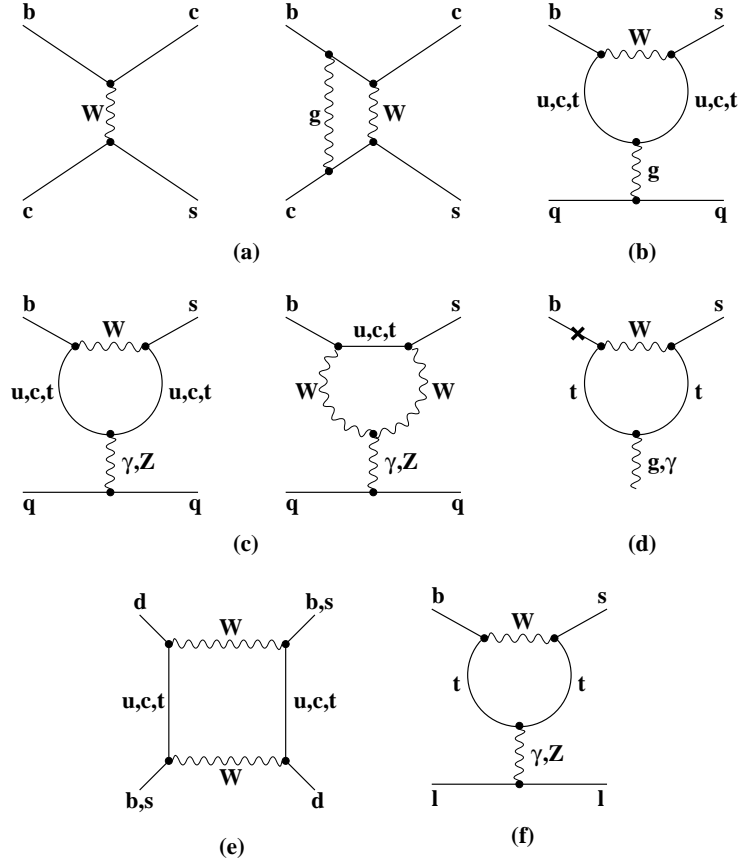}
}
\caption{Typical Penguin and Box Diagrams.}
\label{fig:fdia}
\end{figure}

{\bf Electroweak--Penguins :}
\begin{equation}\label{OS4} 
Q_7 = {3\over 2}\;(\bar s b)_{V-A}\sum_{q=u,d,s,c,b}e_q\;(\bar qq)_{V+A} 
~~~~~ Q_8 = {3\over2}\;(\bar s_{\alpha} b_{\beta})_{V-A}
\sum_{q=u,d,s,c,b}e_q
        (\bar q_{\beta} q_{\alpha})_{V+A}
\end{equation}
\begin{equation}\label{OS5} 
 Q_9 = {3\over 2}\;(\bar s b)_{V-A}\sum_{q=u,d,s,c,b}e_q(\bar q q)_{V-A}
~~~~~Q_{10} ={3\over 2}\;
(\bar s_{\alpha} b_{\beta})_{V-A}\sum_{q=u,d,s,c,b}e_q\;
       (\bar q_{\beta}q_{\alpha})_{V-A} \,.
\end{equation}
Here, $\alpha$ and $\beta$ are colour indices and $e_q$ denotes the 
electrical quark charges reflecting the
electroweak origin of $Q_7,\ldots,Q_{10}$. 
$Q_{2}$, $Q_{3-6}$ and $Q_{7,9}$ originate in the tree level 
$W^\pm$-exchange,
gluon penguin and ($\gamma, Z^0$)-penguin diagrams respectively.
These are the diagrams a)--c) in fig.\, \ref{fig:fdia}.
To generate $Q_1$, $Q_8$ and $Q_{10}$ additional gluonic exchanges
are needed. The operators given above have dimension six. Of interest
are also operators of dimension five which are responsible for the
$B\to s\gamma$ decay. They originate in the diagram d) in 
fig.\, \ref{fig:fdia} where $\gamma$ and the gluon are on-shell. 
They will be given in Section 7. 
In what follows we will neglect the higher dimensional
operators as their contributions to weak decays are marginal.

Now what about the couplings $C_i(\mu)$ and the scale $\mu$? 
The 
important point is that $C_i(\mu)$
summarize the physics contributions from scales higher than $\mu$ and due to
asymptotic freedom of QCD they can be calculated in perturbation theory as
long as $\mu$ is not too small. $C_i$ include the top quark contributions and
contributions from other heavy particles such as W, Z-bosons and charged
Higgs particles or supersymmetric particles in the supersymmetric extensions
of the Standard Model. At higher orders in the electroweak coupling the
neutral Higgs may also contribute. Consequently $C_i(\mu)$ depend generally 
on $m_t$ and also on the masses of new particles if extensions of the 
Standard Model are considered. This dependence can be found by evaluating 
the {\it box} and {\it penguin} diagrams with full W-, Z-, top- and 
new particles exchanges shown in fig. \ref{fig:fdia} and {\it properly} 
including short distance QCD 
effects. The latter govern the $\mu$-dependence of the couplings $C_i(\mu)$.

The value of $\mu$ can be chosen arbitrarily. It serves 
to separate the physics contributions to a given decay amplitude into
short-distance contributions at scales higher than $\mu$ and long-distance
contributions corresponding to scales lower than $\mu$. It is customary 
to choose
$\mu$ to be of the order of the mass of the decaying hadron. 
This is $\ord (\mb)$ and $\ord(\mc)$ for B-decays and
D-decays respectively. In the case of K-decays the typical choice is
 $\mu=\ord(1-2~GeV)$
instead of $\ord(m_K)$, which is much too low for any perturbative 
calculation of the couplings $C_i$.

Now due to the fact that $\mu\ll  M_{W,Z},~ m_t$, large logarithms 
$\ln\mw/\mu$ compensate in the evaluation of
$C_i(\mu)$ the smallness of the QCD coupling constant $\alpha_s$ and 
terms $\alpha^n_s (\ln\mw/\mu)^n$, $\alpha^n_s (\ln\mw/\mu)^{n-1}$ 
etc. have to be resummed to all orders in $\alpha_s$ before a reliable 
result for $C_i$ can be obtained.
This can be done very efficiently by means of the renormalization group
methods \cite{REGM,HV1,Weinberg}. 
Indeed solving the renormalization group equations for the
Wilson coefficients $C_i(\mu)$ summs automatically large
logarithms.
The resulting {\it renormalization group improved} perturbative
expansion for $C_i(\mu)$ in terms of the effective coupling constant 
$\alpha_s(\mu)$ does not involve large logarithms and is more reliable.

It should be stressed at this point that the construction of the effective
Hamiltonian ${\cal H}_{eff}$ by means of the operator product expansion and 
the
renormalization group methods can be done fully in the perturbative framework.
The fact that the decaying hadrons are bound states of quarks is irrelevant
for this construction. Consequently the coefficients $C_i(\mu)$ are 
independent of the
particular decay considered in the same manner in which the usual gauge
couplings are universal and process independent.

Having constructed the effective Hamiltonian we can proceed
to evaluate the decay amplitudes. An amplitude for a decay of a given meson 
$M= K, B,..$ into a final state $F=\pi\nu\bar\nu,~\pi\pi,~DK$ is simply 
given by
\be\label{amp5}
A(M\to F)=\langle F|{\cal H}_{eff}|M\rangle
=\frac{G_F}{\sqrt{2}}\sum_i V^i_{CKM} C_i(\mu)\langle F|Q_i(\mu)|M\rangle,
\ee
where $\langle F|Q_i(\mu)|M\rangle$ 
are the hadronic matrix elements of $Q_i$ between M and F. As indicated
in (\ref{amp5}) these matrix elements depend similarly to $C_i(\mu)$ 
on $\mu$. They summarize the physics contributions to the amplitude 
$A(M\to F)$ from scales lower than $\mu$.

We realize now the essential virtue of OPE: it allows to separate the problem
of calculating the amplitude
$A(M\to F)$ into two distinct parts: the {\it short distance}
(perturbative) calculation of the couplings $C_i(\mu)$ and 
the {\it long-distance} (generally non-perturbative) calculation of 
the matrix elements $\langle Q_i(\mu)\rangle$. The scale $\mu$, as
advertised above, separates then the physics contributions into short
distance contributions contained in $C_i(\mu)$ and the long distance 
contributions
contained in $\langle Q_i(\mu)\rangle$. By evolving this scale from 
$\mu=\ord(\mw)$ down to lower values one
simply transforms the physics contributions at scales higher than $\mu$ 
from the hadronic matrix elements into $C_i(\mu)$. Since no information 
is lost this way the full amplitude cannot depend on $\mu$. Therefore 
the $\mu$-dependence of the couplings $C_i(\mu)$ has to cancel the 
$\mu$-dependence of $\langle Q_i(\mu)\rangle$. In other words it is a
matter of choice what exactly belongs to $C_i(\mu)$ and what to 
$\langle Q_i(\mu)\rangle$. This cancellation
of $\mu$-dependence involves generally several terms in the expansion 
in (\ref{amp5}).

Clearly, in order to calculate the amplitude $A(M\to F)$, the matrix 
elements $\langle Q_i(\mu)\rangle$ have to be evaluated. 
Since they involve long distance contributions one is forced in
this case to use non-perturbative methods such as lattice calculations, the
1/N expansion (N is the number of colours), QCD sum rules, hadronic sum rules,
chiral perturbation theory and so on. In the case of certain B-meson decays,
the {\it Heavy Quark Effective Theory} (HQET) turns out to be a 
useful tool.
Needless to say, all these non-perturbative methods have some limitations.
Consequently the dominant theoretical uncertainties in the decay amplitudes
reside in the matrix elements $\langle Q_i(\mu)\rangle$.

The fact that in most cases the matrix elements $\langle Q_i(\mu)\rangle$
 cannot be reliably
calculated at present, is very unfortunate. One of the main goals of the
experimental studies of weak decays is the determination of the CKM factors 
$V^i_{\rm CKM}$
and the search for the physics beyond the Standard Model. Without a reliable
estimate of $\langle Q_i(\mu)\rangle$ this goal cannot be achieved unless 
these matrix elements can be determined experimentally or removed from the 
final measurable quantities
by taking the ratios or suitable combinations of amplitudes or branching
ratios. However, this can be achieved only in a handful of decays and
generally one has to face directly the calculation of 
$\langle Q_i(\mu)\rangle$.

Now in the case of semi-leptonic decays, in which there is at most one hadron
in the final state, the chiral perturbation theory in the case of K-decays
and HQET in the case of B-decays have already provided useful estimates of
the relevant matrix elements. This way it was possible to achieve
satisfactory determinations of the CKM elements $V_{us}$ and $V_{cb}$ in 
$K\to\pi e\nu$ and $B\to D^*e\nu$ respectively. 
Similarly certain rare decays like $K\to\pi\nu\bar\nu$ and
$B\to\mu\bar\mu$ can be calculated very reliably.

The case of non-leptonic decays in which the final state consists exclusively
out of hadrons is a completely different story. Here even the matrix
elements entering the simplest decays, the two-body decays like 
$K\to\pi\pi$, $D\to K\pi$ or $B\to DK$ cannot be
calculated in QCD reliably at present. 
More promising in this respect is the evaluation of hadronic matrix
elements relevant for $K^0-\bar K^0$ and $B^0_{d,s}-\bar B^0_{d,s}$
mixings.

Returning to the Wilson coefficients $C_i(\mu)$ it should be stressed that 
similar
to the effective coupling constants they do not depend only on the scale $\mu$
but also on the renormalization scheme used: this time on the 
scheme for the renormalization of local operators. That the local operators 
undergo renormalization is not surprising. After all they represent effective
vertices and as the usual vertices in a field theory they have to be
renormalized when quantum corrections like QCD or QED corrections are taken
into account. As a consequence of this, the hadronic matrix elements 
$\langle Q_i(\mu)\rangle$
are
renormalization scheme dependent and this scheme dependence must be cancelled
by the one of $C_i(\mu)$ so that the physical amplitudes are 
renormalization scheme
independent. Again, as in the case of the $\mu$-dependence, the 
cancellation of
the renormalization scheme dependence involves generally several 
terms in the
expansion (\ref{amp5}).

Now the $\mu$ and the renormalization scheme dependences of the couplings 
$C_i(\mu)$ can
be evaluated efficiently in the renormalization group improved perturbation
theory. Unfortunately the incorporation of these dependences in the
non-perturbative evaluation of the matrix elements  
$\langle Q_i(\mu)\rangle$
remains as an important
challenge and most of the non-perturbative methods on the market are
insensitive to these dependences. The consequence of this unfortunate
situation is obvious: the resulting decay amplitudes are $\mu$ and 
renormalization
scheme dependent which introduces potential theoretical uncertainty in the
predictions. On the other hand 
in certain decays these dependences can be put under control.

So far I have discussed only  {\it exclusive} decays. It turns out that
in the case of {\it inclusive} decays of heavy mesons, like B-mesons,
things turn out to be easier. In an inclusive decay one sums over all 
(or over
a special class) of accessible final states so that the amplitude for an
inclusive decay takes the form:
\be\label{ampi}
A(B\to X)
=\frac{G_F}{\sqrt{2}}\sum_{f\in X}V^i_{\rm CKM} 
C_i(\mu)\langle f|Q_i(\mu)|B\rangle~.
\ee
At first sight things look as complicated as in the case of exclusive decays.
It turns out, however, that the resulting branching ratio can be calculated
in the expansion in inverse powers of $\mb$ with the leading term 
described by the spectator model
in which the B-meson decay is modelled by the decay of the $b$-quark:
\be\label{hqe}
{\rm Br}(B\to X)={\rm Br}(b\to q) +\ord(\frac{1}{\mb^2})~. 
\ee
This formula is known under the name of the Heavy Quark Expansion (HQE)
\cite{HQE1}-\cite{HQE3}.
Since the leading term in this expansion represents the decay of the quark,
it can be calculated in perturbation theory or more correctly in the
renormalization group improved perturbation theory. It should be realized
that also here the basic starting point is the effective Hamiltonian 
 (\ref{b1})
and that the knowledge of the couplings $C_i(\mu)$ is essential for 
the evaluation of
the leading term in (\ref{hqe}). But there is an important difference 
relative to the
exclusive case: the matrix elements of the operators $Q_i$ can be 
"effectively"
evaluated in perturbation theory. 
This means, in particular, that their $\mu$ and renormalization scheme
dependences can be evaluated and the cancellation of these dependences by
those present in $C_i(\mu)$ can be explicitly investigated.

Clearly in order to complete the evaluation of $Br(B\to X)$ also the 
remaining terms in
(\ref{hqe}) have to be considered. These terms are of a non-perturbative 
origin, but
fortunately they are suppressed by at least two powers of $m_b$. 
They have been
studied by several authors in the literature with the result that they affect
various branching ratios by less than $10\%$ and often by only a few percent.
Consequently the inclusive decays give generally more precise theoretical
predictions at present than the exclusive decays. On the other hand their
measurements are harder. There is of course an important theoretical
issue related to the validity of HQE in (\ref{hqe}) which appear in the 
literature under the name of quark-hadron duality. 
I will not discuss it here. Recent discussions of this issue 
can be found in  \cite{DUALITY}.

We have learned now that the matrix elements of $Q_i$ are easier to handle in
inclusive decays than in the exclusive ones. On the other hand the evaluation
of the couplings $C_i(\mu)$ is equally  difficult in both cases although 
as stated
above it can be done in a perturbative framework. Still in order to achieve
sufficient precision for the theoretical predictions it is desirable to have
accurate values of these couplings. Indeed it has been realized at the end of
the eighties
that the leading term (LO) in the renormalization group improved perturbation
theory, in which the terms $\alpha^n_s (\ln\mw/\mu)^n$ are summed, is 
generally insufficient and the
inclusion of next-to-leading corrections  (NLO) which correspond to summing
the terms $\alpha^n_s (\ln\mw/\mu)^{n-1}$ is necessary. 
In particular, unphysical left-over $\mu$-dependences
in the decay amplitudes and branching ratios resulting from the truncation of
the perturbative series are considerably reduced by including NLO
corrections. These corrections are known by now for the most important and
interesting decays and will be briefly reviewed below.

\section{Penguin--Box Expansion and OPE}
The FCNC decays, in particular rare and CP violating decays are governed 
by various penguin and box diagrams with internal top quark and charm quark
exchanges. Some examples are shown in fig. 2. 
These diagrams can be evaluated in the full theory and are summarized
by
a set of basic universal (process independent) 
$\mt$-dependent functions $F_r(x_t)$ \cite{IL} where $x_t=\mt^2/\mw^2$. 
Explicit expressions for these
functions can be found in \cite{BBL,BF97,AJB98}. 

It is useful to express the OPE formula (\ref{amp5}) directly in terms
of the functions $F_r(x_t)$ \cite{PBE0}.
To this end we  
rewrite the $A(M\to F)$ in (\ref{amp5}) as follows
\begin{equation}\label{USE}
A(M\to F)=\frac{G_{\rm F}}{\sqrt 2} V_{\rm CKM}
\sum_{i,k} \langle F\mid O_k(\mu)\mid M\rangle \;\hat U_{ki}\;(\mu,\mw) 
          \; C_i(\mw),
\end{equation}
where $\hat U_{kj}(\mu,M_W)$ is the renormalization group
transformation from $\mw$ down to $\mu$. Explicit formula for this
transformation will be given below. In order to simplify the presentation
we have removed the index ``i" from $V^i_{\rm CKM}$

Now
$C_i(\mw)$ are linear combinations of the basic functions
$F_r(x_t)$ so that we can write
\be\label{CIA}
C_i(\mw)=c_i+\sum_r h_{ir} F_r(x_t)
\ee
where $c_i$ and $h_{ir}$ are $\mt$-independent constants. 
Inserting (\ref{CIA})
into (\ref{USE}) and summing over $i$ and $k$ we 
find
\begin{equation}
A({M\to F}) = P_0(M\to F) + \sum_r P_r(M\to F) \, F_r(x_t),
\label{generalPBE1}
\end{equation}
 with
\be\label{PBE8}
P_0(M\to F)= \sum_{i,k} \langle F\mid O_k(\mu)\mid M\rangle 
\;\hat U_{ki}\;(\mu,\mw)c_i~,
\ee 
\be\label{PBE9}
P_r(M\to F)= \sum_{i,k} \langle F\mid O_k(\mu)\mid M\rangle 
\;\hat U_{ki}\;(\mu,\mw)h_{ir}~,
\ee 
where we have suppressed the overall factor $(G_F/\sqrt{2})V_{CKM}$.
I would like to call (\ref{generalPBE1}) {\it Penguin-Box Expansion} 
(PBE) \cite{PBE0}.

The coefficients $P_0$ and $P_r$ are process dependent. 
This process dependence  enters through
$\langle F\mid O_k(\mu)\mid M\rangle$. In certain cases like
$K\to\pi\nu\bar\nu$ these matrix elements are very simple implying
simple formulae for the coefficients $P_0$ and $P_r$. In other
situations, like $\epe$, this is not the
case.

Originally  PBE was designed to expose the $\mt$-dependence
of FCNC processes \cite{PBE0}. After the top quark mass has been
measured precisely this role of PBE is less important.
On the other hand,
PBE is very well suited for the study of the extentions of the
Standard Model in which new particles are exchanged in the loops.
We know already that these particles are heavier than W-bosons
and consequently they can be integrated out together with
the weak bosons and the top quark. If there are no new local operators
the mere change is to modify the functions $F_r(x_t)$ which now
acquire the dependence on the masses of new particles such as
charged Higgs particles and supersymmetric particles. The process
dependent coefficients $P_0$ and $P_r$ remain unchanged. This
is particularly useful as the most difficult part is the evaluation
of $\hat U_{kj}(\mu,M_W)$ and of the hadronic matrix elements, both
contained in these coefficients. 
However, if
new effective operators with different Dirac and colour structures
are present the values of $P_0$ and $P_r$ are modified. 
Examples of the applications of PBE to physics
beyond the Standard Model can be found in \cite{BBHLS,MW96,AAA}.

The universality of the functions $F_r(x_t)$ 
can be violated partly when QCD corrections to one loop penguin
and box diagrams are included. For instance in the case of
semi-leptonic FCNC transitions there is no gluon exchange in
a $Z^0$-penguin diagram parallel to the $Z^0$-propagator but
such an exchange takes place in non-leptonic decays in which the
bottom line is a quark-line. Thus the general universality of $F_r(x_t)$
present at one loop level is reduced to two universality classes
relevant for semi-leptonic and non-leptonic transitions.
However, 
the $\ord(\as)$ corrections to the functions $F_r(x_t)$ are
generally rather small when the top quark mass $\mtb(\mt)$
is used and consequently
the inclusion of QCD effects plays mainly the role in reducing 
various  $\mu$-dependences.

In order to see the general structure of $A(M\to F)$ more transparently
let us write it as follows:
\be\label{GS}
A(M\to F) =B_{M\to F} V_{\rm CKM} \eta_{\rm QCD} F(x_t) +{\rm Charm}
\ee
where the first term represents the internal top quark contribution
and "Charm" stands for remaining contributions, in particular those
with internal charm quark exchanges. $F(x_t)$ represents one of the
universal functions and $\eta_{\rm QCD}$ the corresponding short
distance QCD corrections. The parameter $B_{M\to F}$ represents
the relevant hadronic matrix element, which can only be calculated by
means of non-perturbative methods. However, in certain lucky situations 
$B_{M\to F}$ can be extracted from well measured leading decays and when it
enters also  other decays, the latter are then free from hadronic 
uncertainties and offer very useful means for extraction of CKM parameters.
One such example is the decay $\kpn$ for which one has

\be\label{KPNN}
Br(\kpn)=\left[\frac{\alpha_{\rm QED}^2 Br(K^+ \to\pi^0 e^+\nu)}
{V^2_{us} 2 \pi^2 \sin^4 \theta_W}\right] \cdot
\left| V^*_{ts} V_{td} \eta_{\rm QCD}^t F(x_t) +
V^*_{cs} V_{cd} \eta_{\rm QCD}^c F(x_c)\right |^2
\ee
 The factor in square brackets stands for the $" B-factor"$
in (\ref{GS}), which is given in terms of well measured quantities. Since
$V_{cs}$, $V_{cd}$ and $V_{ts}$ are already rather well
determined and $F(x_i)$ and $\eta_{\rm QCD}^i$ can be calculated
in perturbation theory, the element $V_{td}$ can be extracted
from $Br(\kpn)$ without essentially any theoretical uncertainties.
We will be more specific about this in Section 7.

\section{Motivations for NLO Calculations}
Going beyond the LO approximation for $C_i(\mu)$ 
is certainly an important but a 
non-trivial step. For this reason one needs some motivations to perform
this step. Here are the main reasons for going beyond LO:
\begin{itemize}
\item The NLO is first of all necessary to test the validity of
the renormalization group improved perturbation theory.
\item Without going to NLO the QCD scale $\Lambda_{\overline{MS}}$
\cite{BBDM} 
extracted from various high energy processes cannot be used 
meaningfully in weak decays.
\item 
Due to renormalization group invariance the physical
amplitudes do not depend on the scales $\mu$ present in $\alpha_s$
or in the running quark masses, in particular $\mt(\mu)$, 
$\mb(\mu)$ and $\mc(\mu)$. However,
in perturbation theory this property is broken through the truncation
of the perturbative series. Consequently one finds sizable scale
ambiguities in the leading order, which can be reduced considerably
by going to NLO. 
\item
The Wilson Coefficients are renormalization scheme dependent quantities.
This scheme dependence appears first at NLO. For a proper matching of
the short distance contributions to the long distance matrix elements
obtained from lattice calculations it is essential to calculate NLO.
The same is true for inclusive heavy quark decays in which the hadron
decay can be modeled by a decay of a heavy quark and the matrix elements
of $Q_i$ can be effectively calculated in an expansion in $1/\mb$.
\item 
In several cases the central issue of the top quark mass dependence
is strictly a NLO effect.
\end{itemize}

\section{General Structure of Wilson Coefficients}
We will give here a formula for the Wilson coefficient $C(\mu)$ of
a single operator $Q$ including NLO corrections. The case of
several operators which mix under renormalization is much more
complicated. Explicit formulae are given in \cite{BBL,AJB98}.

$C(\mu)$
is given by
\begin{equation}\label{C1+-}
 C(\mu) = U(\mu,\mw) C(\mw) 
  \end{equation}
where 
\begin{equation}\label{U1+-}
U(\mu,\mw)= \exp \left[ 
  \int_{g_s(\mw)}^{g_s(\mu)}{dg' \frac{\gamma_Q(g_s')}{\beta(g_s')}}\right] 
\end{equation}
is the evolution function, which allows to calculate
$C(\mu)$ once $C(\mw)$ is known. The latter can be calculated in
perturbation theory in the process of integrating out $W^\pm$, $Z^0$
and top quark fields. Details can be found in \cite{BBL,AJB98}.
Next $\gamma_Q$ is the anomalous dimension of the operator Q and
$\beta(g_s)$ is the renormalization group function which governs
the evolution of the QCD coupling constant $\alpha_s(\mu)$.

At NLO we have
\begin{equation}\label{B8}
C(M_W)=1+\frac{\as(M_W)}{4\pi}B
\end{equation}

\begin{equation}\label{gg01P}
\gamma_Q(\as)=\gamma_Q^{(0)}\aspi + \gamma_Q^{(1)}\left(\aspi\right)^2
\end{equation}

\begin{equation}\label{bg01P}
\beta(g_s)=-\beta_0{g_s^3\over 16\pi^2}-\beta_1{g_s^5\over (16\pi^2)^2}
  \end{equation}

Inserting the last two formulae into (\ref{U1+-}) and expanding in
$\alpha_s$ we find
\begin{equation}\label{B9P}
U(\mu,\mw)=\left[1+\frac{\as(\mu)}{4\pi}J\right]
      \left[\frac{\as(M_W)}{\as(\mu)}\right]^{d}
\left[1-\frac{\as(M_W)}{4\pi} J\right]
\end{equation}
with
\begin{equation}\label{B10P}
J=\frac{d}{\beta_0}\beta_1-\frac{\gamma^{(1)}_Q}{2\beta_0}
\qquad\qquad
d=\frac{\gamma^{(0)}_Q}{2\beta_0}.
\end{equation}

Inserting (\ref{B9P}) and (\ref{B8}) into (\ref{C1+-}) we find
an important formula for $C(\mu)$ in the NLO approximation:
\begin{equation}\label{B9PP}
C(\mu)=\left[1+\frac{\as(\mu)}{4\pi}J\right]
      \left[\frac{\as(M_W)}{\as(\mu)}\right]^{d}
\left[1+\frac{\as(M_W)}{4\pi}(B-J)\right]~.
\end{equation}

\begin{table}[thb]
\caption{References to NLO Calculations}
\label{TAB1}
\begin{center}
\begin{tabular}{|l|l|}
\hline
\bf \phantom{XXXXXXXX} Decay & \bf Reference \\
\hline
\hline
\multicolumn{2}{|c|}{$\Delta F=1$ Decays} \\
\hline
current-current operators     & \cite{ACMP,WEISZ} \\
QCD penguin operators         & \cite{BJLW1,BJLW,ROMA1,ROMA2,MIS1} \\
electroweak penguin operators & \cite{BJLW2,BJLW,ROMA1,ROMA2} \\
magnetic penguin operators    & \cite{MisMu:94,CZMM} \\
$Br(B)_{SL}$                  & \cite{ACMP,Buch:93,Bagan,LUO} \\
inclusive $\Delta S=1$ decays       & \cite{JP} \\
\hline
\multicolumn{2}{|c|}{Particle-Antiparticle Mixing} \\
\hline
$\eta_1$                   & \cite{HNa} \\
$\eta_2,~\eta_B$           & \cite{BJW90,UKJS} \\
$\eta_3$                   & \cite{HNb} \\
\hline
\multicolumn{2}{|c|}{Rare $K$- and $B$-Meson Decays} \\
\hline
$K^0_L \rightarrow \pi^0\nu\bar{\nu}$, $B \rightarrow l^+l^-$,
$B \rightarrow X_{\rm s}\nu\bar{\nu}$ & \cite{BB1,BB2,MIUR,BB98} \\
$K^+   \rightarrow \pi^+\nu\bar{\nu}$, $K_{\rm L} \rightarrow \mu^+\mu^-$
                                      & \cite{BB3,BB98} \\
$K^+\to\pi^+\mu\bar\mu$               & \cite{BB5} \\
$K_{\rm L} \rightarrow \pi^0e^+e^-$         & \cite{BLMM} \\
$B\rightarrow X_s \mu^+\mu^-$           & \cite{Mis:94,BuMu:94} \\
$B\rightarrow X_s \gamma$      & 
\cite{AG2}-\cite{BG98} \\
$\Delta\Gamma_{B_s}$     &  \cite{BBGLN} \\
inclusive B $\to$ Charmonium & \cite{BMR} \\
\hline
\multicolumn{2}{|c|}{Two-Higgs Doublet Models} \\
\hline
$B\rightarrow X_s \gamma$  & \cite{GAMB,strum,BG98} \\
\hline
\multicolumn{2}{|c|}{Supersymmetry} \\
\hline
$\Delta M_K$ and $\varepsilon_K$ & \cite{CET0,CET} \\
$B\rightarrow X_s \gamma$ & \cite{GAMB2} \\
\hline
\end{tabular}
\end{center}
\end{table}

\section{Status of NLO Calculations}
Since the pioneering leading order calculations of Wilson coefficients
for current--current \cite{MKG,ALMA} and penguin operators \cite{PENGUIN},
enormous progress has been made, so that at present most of the decay
amplitudes are known at the NLO level.
We list all 
existing NLO calculations for weak decays in table \ref{TAB1}.
In addition to the calculations in the Standard Model we list the
calculations in two-Higgs doublet models and supersymmetry.
In table \ref{TAB2} we list references to calculations of two-loop
electroweak contributions to rare decays. The latter calculations
allow to reduce scheme and scale dependences related to the
definition of electroweak parameters like $\sin^2\theta_W$,
$\alpha_{QED}$, etc. Next, useful techniques for three-loop
calculations can be found in \cite{MIS2} and a very general
discussion of the evanescent operators including earlier references
is presented in \cite{HNE}.
Further details on these calculations can be found in the orignal
papers, in the review \cite{BBL} and in the Les Houches lectures
\cite{AJB98}. Some of the implications
of these calculations will be analyzed briefly in subsequent
sections.

\begin{table}[bth]
\caption{Electroweak Two-Loop Calculations}
\label{TAB2}
\begin{center}
\begin{tabular}{|l|l|}
\hline
\bf \phantom{XXXXXXXX} Decay & \bf Reference \\
\hline
$K^0_L \rightarrow \pi^0\nu\bar{\nu}$, $B \rightarrow l^+l^-$,
$B \rightarrow X_{\rm s}\nu\bar{\nu}$ & \cite{BB97} \\
$B\rightarrow X_s \gamma$ & \cite{CZMA,STRUMIA,KN98} \\
$B^0-\bar B^0$ mixing & \cite{GKP} \\
\hline
\end{tabular}
\end{center}
\end{table}

\section{Applications: News}
\subsection{Preliminaries}
There is a vast literature on the applications of NLO calculations
listed in table 1. As they are already reviewed in detail in 
\cite{BBL,BF97,AJB98} there is no point to review them here again.
I will rather discuss briefly some of the most important applications
in general terms. This will also give me the opportunity to update  
some of the numerical results presented in \cite{AJB98}. This update
is related mainly to the improved experimental lower bound on 
$B_s^0-\bar B_s^0$ mixing $((\Delta M)_s> 12.4/ps)$ and a slight increase
in $|V_{ub}|/|V_{cb}|$: $0.091\pm0.016$, both presented at the last
Rochester Conference in Vancouver \cite{RUDO}.

\begin{figure}[hbt]
\vspace{0.10in}
\centerline{
\epsfysize=2.1in
\epsffile{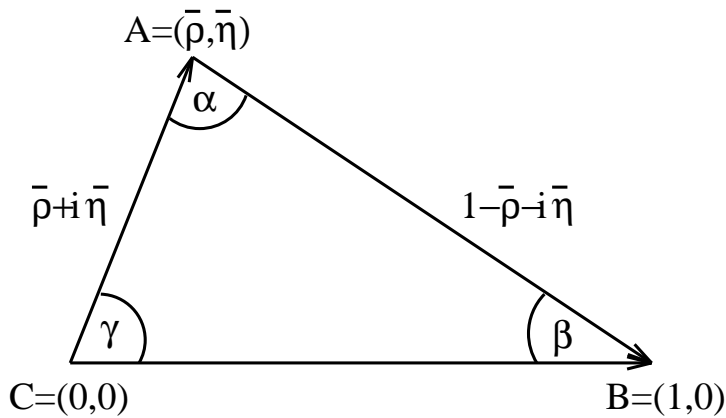}
}
\vspace{0.08in}
\caption{Unitarity Triangle.}\label{fig:utriangle}
\end{figure}

\subsection{Unitarity Triangle}
The standard analysis of the unitarity triangle 
(see fig. \ref{fig:utriangle}) uses
the values of $|V_{us}|$, $\vcb$, $\vub$ extracted from tree level
K- and B- decays, the indirect CP-violation in $K_L\to\pi\pi$
represented by the parameter $\varepsilon$ and the 
$B^0_{d,s}-\bar B^0_{d,s}$ mixings described by the mass differences
$(\Delta M)_{d,s}$. From this analysis follows the allowed range for 
$(\bar\varrho,\bar\eta)$ describing the apex of the unitarity
triangle. Here \cite{BLO}
\begin{equation}\label{2.88d}
\bar\varrho=\varrho (1-\frac{\lambda^2}{2}),
\qquad
\bar\eta=\eta (1-\frac{\lambda^2}{2}).
\end{equation}
where $\lambda$, $\varrho$ and $\eta$ are Wolfenstein parameters \cite{WOL}
with $|V_{us}|=\lambda=0.22$. We have in particular
\begin{equation}\label{CKM2}
V_{ub}=\lambda\vcb(\varrho-i\eta),
\qquad
V_{td}=\lambda\vcb(1-\bar\varrho-i\bar\eta).
\end{equation}
$\eta\not=0$ is responsible for CP violation in the Standard Model.

\begin{figure}[thb]
\vspace{0.10in}
\centerline{
\epsfysize=3.4in
\rotate[r]{\epsffile{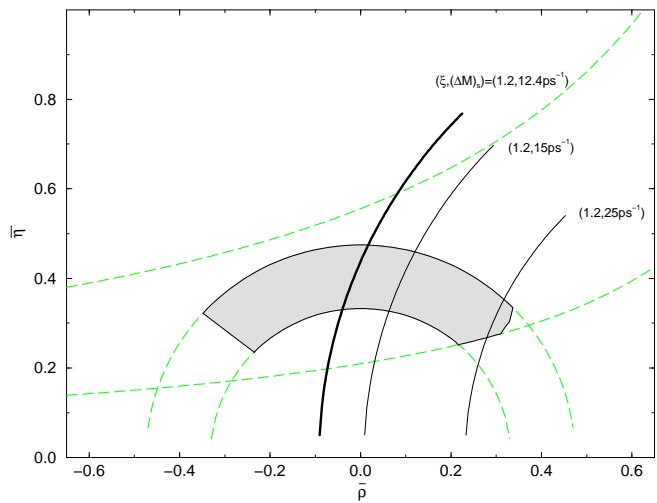}}
}
\vspace{0.08in}
\caption[]{
Unitarity Triangle 1998.
\label{fig:utdata}}
\end{figure}

The allowed region for $(\bar\varrho,\bar\eta)$ is presented in fig. 
\ref{fig:utdata}.
It is the shaded area on the right hand side of the solid circle
which represents 
the upper bound for $(\Delta M)_d/(\Delta M)_s$. The hyperbolas
give the constraint from $\varepsilon$ and the two circles centered
at $(0,0)$ the constraint from $\vub$. The white areas between the
lower $\varepsilon$-hyperbola and the shaded region are excluded
by $B^0_{d}-\bar B^0_{d}$ mixing. We observe that the region
$\bar\varrho<0$ is practically excluded. The main remaining theoretical
uncertainties in this analysis are the values of non-perturbative
parameters: $B_K$ in $\varepsilon$, $F_{B_d}\sqrt{B_d}$ in
$(\Delta M)_d$ and $\xi=F_{B_s}\sqrt{B_s}/F_{B_d}\sqrt{B_d}$
in $(\Delta M)_d/(\Delta M)_s$. I have used $B_K=0.80\pm 0.15$,
$F_{B_d}\sqrt{B_d}= 200\pm 40 \mev$ and $\xi_{\rm max}=1.2$. On the
experimental side $\vub$ and $(\Delta M)_s$ should be improved.

From this analysis we extract
\be
\vtd =(8.6\pm1.6)\cdot 10^{-3}, \qquad
\IM(V^*_{ts} V_{td})=(1.38\pm 0.33)\cdot 10^{-4}
\ee
and
\be
\sin 2\beta =0.71\pm 0.13, \qquad \sin\gamma=0.83\pm 0.17
\ee

\subsection{ $\varepsilon'/\varepsilon$}

 $\varepsilon'/\varepsilon$ is the ratio of the direct and indirect
CP violation in $K_L\to\pi\pi$. A measurement of a non-vanishing value of 
$\epe$ would give the first
signal for the direct CP violation ruling out the superweak models
\cite{wolfenstein:64}.
In the Standard
Model $\varepsilon'/\varepsilon $ is governed by QCD penguins and
electroweak (EW) penguins. The corresponding operators are given in
(\ref{OS2})-(\ref{OS5}).
With increasing $\mt$ the EW penguins become
increasingly important \cite{flynn:89,buchallaetal:90}, and entering
$\varepsilon'/\varepsilon$ with the opposite sign to QCD penguins
suppress this ratio for large $\mt$. For $\mt\approx 200\,\gev$ the ratio
can even be zero \cite{buchallaetal:90}.  Because of this strong
cancellation between two dominant contributions and due to uncertainties
related to hadronic matrix elements of the relevant local operators, a
precise prediction of $\varepsilon'/\varepsilon$ is not possible at
present. 

A very simplified formula (not to be used
for any serious numerical analysis) which
exhibits main uncertainties 
is given as follows
\begin{equation}\label{7e}
\frac{\varepsilon'}{\varepsilon}=15\cdot 10^{-4}\left[ 
\frac{\eta\lambda \vcb^2}{1.3\cdot 10^{-4}}\right]
\left[\frac{120~MeV}{ m_s(2~\gev)} \right]^2 
\left[\frac{\Lms^{(4)}}{300~\mev} \right]^{0.8} 
[B_6-Z(x_t)B_8]
\end{equation} 
where $Z(x_t)\approx 0.18 (\mt/\mw)^{1.86}$ represents the
leading $\mt$-dependence of EW penguins. 
$B_6$ and $B_8$ represent the hadronic matrix elements of the dominant
QCD-penguin operator $Q_6$ and the dominant electroweak penguin
operator $Q_8$ (see (\ref{OS3}) and (\ref{OS4})) respectively.
Together with $\ms(2\gev)$ the values of 
these parameters constitute the main theoretical
uncertainty in evaluating $\epe$. Present status of $\ms$, $B_6$ and
$B_8$ is reviewed in \cite{gupta,AJB98}. Roughly one has $B_6=1.0 \pm 0.2$
and $B_8=0.7\pm 0.2$. Taking these values, $\eta$ of 
fig. \ref{fig:utdata}  and 
$\vcb=0.040\pm 0.003$, I find:

\begin{equation}\label{eprimef}
\varepsilon'/\varepsilon =\left\{ \begin{array}{ll}
(5.7 \pm 3.6)\cdot 10^{-4}~, &~~\ms(2 \gev)=130\pm20\mev \\
(9.1  \pm 5.7)\cdot 10^{-4}~, & ~~\ms(2 \gev)=110\pm20\mev .
\end{array} \right.
\end{equation}
where the chosen values for $\ms$ are in the ball park of various QCD
sum rules and lattice estimates \cite{gupta}.
This should be compared
with the result of NA31 collaboration at CERN which finds
$(\varepsilon'/\varepsilon) = (23 \pm 7)\cdot 10^{-4}$ \cite{barr:93}
and the value of E731 at Fermilab,
$(\varepsilon'/\varepsilon) = (7.4 \pm 5.9)\cdot 10^{-4}$
\cite{gibbons:93}.

The Standard Model expectations are closer to the Fermilab result,
but due to large theoretical and experimental errors no firm conclusion
can be reached at present. The new improved data from CERN and Fermilab
in 1999 and later from DA$\Phi$NE should shed more light on $\epe$.
In this context improved estimates of $B_6$, $B_8$ and $\ms$ are
clearly desirable.

\subsection{$B \to X_s \gamma$}
A lot of efforts have been put into predicting the branching
ratio for the inclusive decay $B \to X_s \gamma$ including NLO
QCD corrections and higher order electroweak corrections. The
relevant references are given in table 1 and in \cite{AJB98},
where details can be found. The final result of these efforts
can be summarized by
\be\label{bsg}
Br(B \to X_s \gamma)_{\rm th}
=( 3.30 \pm 0.15 ({\rm scale})\pm 0.26 ({\rm par}))
                     \cdot 10^{-4}
\ee
where the first error represents residual scale dependences and
the second error is due to uncertainties in input parameters.
The main achievement is the reduction of the scale dependence
through NLO calculations, in particular those given in 
\cite{GREUB} and \cite{CZMM}. In
the leading order the corresponding error would be roughly
$\pm 0.6$ \cite{AG1,BMMP:94}.

The theoretical result in (\ref{bsg})
 should be compared with experimental data:
\begin{equation}\label{bsgexp}
 Br(B \to X_s \gamma)_{\rm exp}=\left\{ \begin{array}{ll}
(3.15 \pm 0.35 \pm 0.41)\cdot 10^{-4}~, &~~{\rm CLEO} \\
(3.11 \pm 0.80 \pm 0.72)\cdot 10^{-4}~, & ~~{\rm ALEPH} ,
\end{array} \right.
\end{equation}
which implies the combined branching ratio:
\begin{equation}\label{bsgex}
 Br(B \to X_s \gamma)_{\rm exp}=
(3.14 \pm 0.48)\cdot 10^{-4}
\end{equation}
Clearly, the Standard Model result agrees well with the data. In
order to see whether any new physics can be seen in this decay,
the theoretical and in particular experimental errors should
be reduced. This is certainly a very difficult task.
Most recent analyses of $B \to X_s\gamma$ in supersymmetric models
and two--Higgs doublet models are listed in table \ref{TAB1}.

\subsection{$K_L\to\pi^0\nu\bar\nu$ and $K^+\to\pi^+\nu\bar\nu$}
$K_L\to\pi^0\nu\bar\nu$ and $K^+\to\pi^+\nu\bar\nu$ are the theoretically
cleanest decays in the field of rare K-decays. 
$K_L\to\pi^0\nu\bar\nu$ is 
dominated by short distance loop diagrams (Z-penguins and box diagrams)
involving the top quark.  $K^+\to\pi^+\nu\bar\nu$ receives
additional sizable contributions from internal charm exchanges.
The great virtue of $K_L\to\pi^0\nu\bar\nu$ is that it proceeds
almost exclusively through direct CP violation \cite{Littenberg} 
and as such is the
cleanest decay to measure this important phenomenon. It also offers
a clean determination of the Wolfenstein parameter $\eta$ and in particular
offers the cleanest measurement
of $ \IM V^*_{ts} V_{td}$ \cite{BB96}. 
$K^+\to\pi^+\nu\bar\nu$ is CP conserving and offers a clean 
determination of $|V_{td}|$. Due to the presence of the charm
contribution and the related $m_c$ dependence it has a small
scale uncertainty absent in $K_L\to\pi^0\nu\bar\nu$.

The next-to-leading QCD corrections 
\cite{BB1,BB2,BB3,MIUR,BB98} to both decays
considerably reduced the theoretical uncertainty
due to the choice of the renormalization scales present in the
leading order expressions, in particular in the charm contribution
to $K^+\to\pi^+\nu\bar\nu$. Since the relevant hadronic matrix
elements of the weak currents entering $K\to \pi\nu\bar\nu$
can be related using isospin symmetry to the leading
decay $K^+ \rightarrow \pi^0 e^+ \nu$, the resulting theoretical
expressions for Br( $K_L\to\pi^0\nu\bar\nu$) and Br($K^+\to\pi^+\nu\bar\nu$)
  are only functions of the CKM parameters, the QCD scale
 $\Lms$
 and the
quark masses $\mt$ and $\mc$. The isospin braking corrections have been
calculated in \cite{MP}.
The long distance contributions to
$K^+ \rightarrow \pi^+ \nu \bar{\nu}$ have been
considered in \cite{RS} and found to be very small: a few percent of the
charm contribution to the amplitude at most, which is safely neglegible.
The long distance contributions to $K_L\to\pi^0\nu\bar\nu$ are negligible
as well \cite{BUSA}.

The explicit expressions for $Br(K^+ \rightarrow \pi^+ \nu \bar{\nu})$ 
and $Br(K_L\to\pi^0\nu\bar\nu)$ can be found in \cite{BBL,BF97,AJB98}. 
Here we
give approximate expressions in order to exhibit various dependences:

\begin{equation}\label{bkpn}
Br(K^+ \rightarrow \pi^+ \nu \bar{\nu})=
0.7\cdot 10^{-10}\left[\left [ \frac{|V_{td}|}{0.010}\right ]^2
\left [\frac{\mid V_{cb}\mid}{0.040} \right ]^2
\left [\frac{\mtb(\mt)}{170~\gev} \right ]^{2.3} 
 +{\rm cc+tc}\right]
\end{equation}

\begin{equation}\label{bklpn}
Br(K_L\to\pi^0\nu\bar\nu)=
3.0\cdot 10^{-11}\left [ \frac{\eta}{0.39}\right ]^2
\left [\frac{\mtb(\mt)}{170~GeV} \right ]^{2.3} 
\left [\frac{\mid V_{cb}\mid}{0.040} \right ]^4 
\end{equation}
where in (\ref{bkpn}) we have shown explicitly only the pure top
contribution.

The impact of NLO calculations is the reduction of
scale uncertainties in $Br(K^+ \rightarrow \pi^+ \nu \bar{\nu})$
from $\pm 23\%$ to $\pm 7\%$.
This corresponds  to the reduction
in the uncertainty in the determination of $|V_{td}|$ from $\pm 14\%$ to 
$\pm 4\%$. The remaining scale uncertainties in $Br(K_L\to\pi^0\nu\bar\nu)$
and in the determination of $\eta$ are fully negligible.

Updating the analysis of \cite{AJB98} one finds \cite{BB98}:
\begin{equation}\label{pre}
Br(K^+ \rightarrow \pi^+ \nu \bar{\nu})=
(8.2\pm 3.2)\cdot 10^{-11}\quad,\quad
Br(K_L\to\pi^0\nu\bar\nu)=(3.1\pm 1.3)\cdot 10^{-11}
\end{equation}
where the errors come dominantly from the uncertainties in the
CKM parameters. 

\begin{figure}[hbt]
\vspace{0.10in}
\centerline{
\epsfysize=2.7in
\epsffile{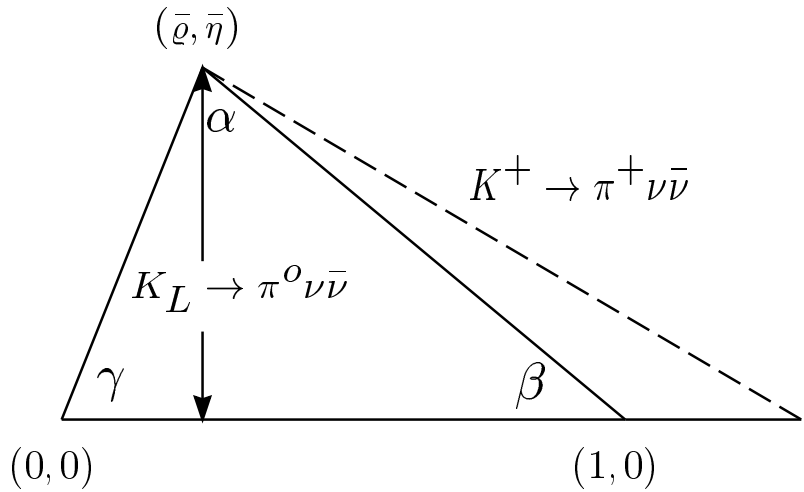}
}
\vspace{0.08in}
\caption{Unitarity triangle from $K\to\pi\nu\bar\nu$.}\label{fig:KPKL}
\end{figure}

 As stressed in \cite{BB96} simultaneous measurements of
$Br(K^+ \rightarrow \pi^+ \nu \bar{\nu})$
and
$Br(K_L\to\pi^0\nu\bar\nu)$
should allow a clean determination of the unitarity triangle
as shown in fig. \ref{fig:KPKL}. In
particular the measurements of these branching ratios with an error
of $\pm 10\%$ will determine $\vtd$, $ \IM V^*_{ts} V_{td}$ and
$\sin 2\beta$ with an accuraccy of $\pm 10\%$, $\pm 5\%$ and
$\pm 0.05$ respectively. The comparision of this determination
of $\sin 2\beta$ with the one by means of the CP-asymmetry in
$B\to \psi K_S$ should offer a very good test of the
Standard Model.

Experimentally we have \cite{kplus}
\begin{equation}
  \label{eq:brkpnnexp}
  BR(\kpn) = (4.2 {+9.7}{-3.5})\cdot 10^{-10}
\end{equation}
and the bound \cite{klong} 
\begin{equation}
  \label{eq:brklexp}
  BR(\klpn) < 1.6 \cdot 10^{-6}.
\end{equation}
Moreover from (\ref{eq:brkpnnexp}) and isospin symmetry one has
\cite{NIR96} $BR(\klpn) < 6.1 \cdot 10^{-9}$. 

The central value in (\ref{eq:brkpnnexp}) is  by a factor of 4 above 
the Standard Model
expectation but in view of large errors the result is compatible with the
Standard Model. 
The analysis of additional
data on $K^+\to \pi^+\nu\bar\nu$ present on tape at BNL787 should narrow
this range in the near future considerably.
In view of the clean character of this decay a measurement of its
branching ratio at the level of $ 2 \cdot 10^{-10}$ 
would signal the presence of physics
beyond the Standard Model \cite{BB98}. The Standard Model sensitivity is
expected to be reached at AGS around the year 2000 \cite{AGS2}.
Also Fermilab with the Main Injector 
could measure this decay \cite{Cooper}.

The present upper bound on $Br(K_{\rm L}\to \pi^0\nu\bar\nu)$ 
is about five orders of magnitude above the Standard Model expectation
(\ref{pre}).
FNAL-E799 expects to reach
the accuracy ${\cal O}(10^{-8})$ and
a very interesting new experiment
at Brookhaven (BNL E926) \cite{AGS2} 
expects to reach the single event sensitivity $2\cdot 10^{-12}$
allowing a $10\%$ measurement of the expected branching ratio. 
There are furthermore plans
to measure this gold-plated  decay with comparable sensitivity
at Fermilab \cite{FNALKL} and KEK \cite{KEKKL}.

\section{Summary}
We have given a general description of OPE and Renormalization
Group techniques as applied to weak decays of mesons. Further
details can be found in \cite{BBL,BF97,AJB98,RF97}. One of the 
outstanding and important challanges for theorists in this field is
a quantitative description of non--leptonic meson decays. In the
field of K--decays this is in particular the case of the
$\Delta I=1/2$ rule for which some progress has been made in
\cite{DI12}. In the field of B--decays progress in a quantitative
description of two--body decays is very desirable in view of
forthcoming B-physics experiments at Cornell, SLAC, KEK, DESY,
FNAL and later at LHC. Recent reviews on non-leptonic two-body
decays are given in \cite{NS97,BUTI,ALI98} 
where further references can be found.

I would like to thank Peter Breitenlohner, Dieter Maison and Julius
Wess for inviting me to such a pleasant symposium.

\end{document}